\documentclass[aps,twocolumn,showpacs,preprintnumbers,amsmath,amssymb,showpacs,showkeys]{revtex4}
\usepackage{slashed}
\usepackage{amssymb}
\usepackage{amsmath}
\usepackage{amsfonts}
\usepackage{epsfig}
\usepackage{graphicx}
\usepackage{tabularx}
\newcommand{\seq}{\begin{subequations}}
\newcommand{\sen}{\end{subequations}}
\newcommand{\eq}{\begin{eqnarray}}
\newcommand{\en}{\end{eqnarray}}

\def\shiftdown#1{#1\llap{\lower.04ex\hbox{#1}}}

\begin{document}

\title{Family of dilatons and metrics for AdS/QCD models} 

\author{Alfredo Vega and Paulina Cabrera}
\vspace*{1.2\baselineskip}

\affiliation{
Instituto de F\'isica y Astronom\'ia y Centro de Astrof\'isica de Valpara\'iso, 
     Universidad de Valpara\'iso,\\
     Avenida Gran Breta\~na 1111, Valpara\'iso, Chile
\vspace*{1.2\baselineskip}\\}

\begin{abstract}

We explore some possibilities for obtaining useful metrics and dilatons 
for AdS/QCD models. As a guideline, we consider dilatons and/or metrics that on the one hand reproduce the mesonic spectrum, and that on the other hand allow us a correct implementation of chiral symmetry breaking in AdS/QCD models. We discuss two procedures: one is based on supersymmetric quantum mechanics techniques and the other considers the interpolation between some limits on dilatons and/or metrics.

\end{abstract}

\date{\today}



\maketitle




\section{Introduction}

The holographic approach to QCD has attracted a great deal of interest, primarily in two directions, i.e., top-down and bottom-up approaches \cite{Erdmenger:2007cm, Brodsky:2014yha, Gursoy:2007cb, Gursoy:2007er}. Top-down approaches start from string theory and, after taking low-energy limits and compacting extra dimensions, attempt to obtain a theory close to QCD. In the bottom - up approach, well-known QCD properties are used as guidelines to build models in a gravity frame with asymptotically anti-de Sitter (AdS) spaces, thereby obtaining models denoted as AdS/QCD.

In the context of bottom-up models, several hadron properties can be easily calculated (see, e.g., \cite{Gutsche:2015xva, Gutsche:2012bp, Gutsche:2012wb, Vega:2010ns, Vega:2012iz, Brodsky:2011xx, Brodsky:2008pf, Brodsky:2007hb, Braga:2011wa, BallonBayona:2007qr, Bellantuono:2015fia, Colangelo:2011xk, BoschiFilho:2012xr}). The first version of an AdS/QCD model is known as hard wall (HW) \cite{Erlich:2005qh, DaRold:2005mxj}, where conformal invariance was broken by introducing a hard cut-off in a holographic coordinate; therefore, a field dual to hadrons cannot propagate in all space. The main problem with HW models is that, because the relationship between $M^{2}$ and $J$ it is not linear, they do not reproduce Regge behavior. To improve this aspect, the soft wall (SW) model was proposed \cite{Karch:2006pv}, where conformal invariance is broken by using a dilaton field. This scalar field allows us to introduce a soft cut-off along the holographic coordinate.

The most common dilaton used is quadratic in a holographic coordinate with AdS metric, and although in this case it is possible to obtain a spectrum with Regge behavior in the mesonic sector, this model is not free of problems. This situation has motivated several authors to search for improvements in this kind of model (see, e.g., \cite{Fang:2016uer, Capossoli:2015ywa, Evans:2015qaa} and references therein). 

In AdS/QCD models the spectrum is calculated using a Schr\"odinger-type equation of motion; therefore, 
using techniques that can achieve strictly isospectral potentials could be very useful when exploring 
different alternatives for dilatons or metrics. This offers us an opportunity to go beyond the typical 
choice for formulating AdS/QCD models.

There are several methods for obtaining a family of strictly isospectral potentials for Schr\"odinger equations (see, e.g.,\cite{AbrahamMoses, Luban:1986iv, Pursey:1986nc, Khare:1989ki, Cooper:1994eh}). In this paper we use one developed in supersymmetric quantum mechanics (SUSY QM) \cite{Cooper:1994eh}. By using SUSY QM techniques we can explore dilatons and/or metrics that reproduce the same mesonic spectrum as in the usual models that consider quadratic dilaton and AdS metrics; additionally, we study the possibility for using SUSY QM techniques in AdS/QCD models that consider chiral symmetry breaking.

Applications of SUSY QM to AdS/QCD models have been 
considered previously \cite{deTeramond:2014asa, Dosch:2015nwa, Dosch:2015bca, Brodsky:2015oia}, though in a different way; these authors consider transformations that produce potentials which are not strictly isospectrals, i.e., in one state the original potential and its supersymmetric partner differ. 
By using these transforms in \cite{deTeramond:2014asa, Dosch:2015nwa, Dosch:2015bca, Brodsky:2015oia} 
the authors have shown that conformal invariance in the equation that describes bound states of quarks 
in light front holography produce interesting superconformal relationships that connect mesonic spectrum with a spectrum of baryons in light and heavy-light sectors.

The idea proposed in this article --- to obtain families of metrics and/or dilatons for SW models --- 
is presented for scalar modes; however, they can be implemented without difficulty in other bosonic or fermionic modes because equations that describe hadrons on the AdS side in SW models generally have the same mathematical structure.

We first obtain a Schr\"odinger-type equation, and from this we derive a family of strictly isospectral potentials, and associated with these new potentials it is possible to extract a family of metrics and/or dilatons to AdS/QCD models. For the sake of simplicity, in this work we studied the family of dilatons related to Schr\"odinger equations that describe scalar glueballs in conventional SW models (i.e., AdS metric and quadratic dilaton) \cite{Colangelo:2007pt,Colangelo:2007if,
BoschiFilho:2012xr}. Because the mass of the gravity mode dual to scalar glueballs in AdS is zero, we obtained simplified equations.

Through the use of SUSY QM techniques, we suppose a family of strictly isospectral potential with a holographical coordinate $z$ that has the same behaviour at small and high values. This means, for example, if we consider as original potential one associated with AdS metric with quadratic dilaton, each dilaton in the family with AdS metric will have 
$\phi (z \rightarrow 0) = \kappa^{2} z^{2}$ and $\phi (z \rightarrow \infty) = \kappa^{2} z^{2}$. By construction, each dilaton in this family produces the same spectrum for scalar glueballs, but unfortunately these dilatons have the same problem as quadratic ones 
at the moment of building AdS/QCD models with chiral symmetry breaking. The problem with chiral symmetry breaking in AdS/QCD models has motivated the study of some modifications to these models \cite{Colangelo:2008us, Gherghetta:2009ac, Kwee:2007nq, Vega:2010ne, Vega:2011tg}.

To solve the problem associated with the correct implementation of chiral symmetry breaking in SW models, some authors have considered dilatons that correspond to an interpolation between $\phi (z \rightarrow 0) = \kappa_{0}^{2} z^{2}$ and $\phi (z \rightarrow \infty) = \kappa_{\infty}^{2} z^{2}$, where $\kappa_{0}^{2} \neq \kappa_{\infty}^{2}$ \cite{Gherghetta:2009ac, Chelabi:2015cwn, Chelabi:2015gpc}; dilatons that interpolate between these limits can be obtained by introducing small changes in the isospectral potential calculated with SUSY QM transforms, as we discuss in this work.

Alternatively, in order to consider a quadratic dilaton in AdS/QCD models with a correct implementation of chiral symmetry breaking, we will show in Sec. IV that metrics in this case must not be asymptotically AdS. This means that a possible modification of SUSY QM equations is complicated, and this procedure is not appropriated for obtaining a family of metrics. Additionally, the use of a warp factor, which is not asymptotically AdS, 
is not permitted within the chosen framework, even though it reproduce almost the same hadron spectrum as the soft-wall model and gives a good description of chiral symmetry breaking.

On the other hand, the authors of \cite{Gherghetta:2009ac} discuss the behaviour of dilatons to small and high $z$, in order to obtain models that allow for the correct incorporation of chiral symmetry breaking in models on AdS space; they suggest dilatons that correspond to interpolation between these limits. In a similar way it is possible to study asymptotic limits for metrics in models with quadratic dilatons. We consider as an example of this procedure one interpolation function for dilatons in AdS spaces, as well as the interpolation for metrics in the case that quadratic dilatons are used.

This article contains six sections. In Sec. II we briefly show how to obtain the spectrum for scalar hadrons with SW models that consider an symptotically AdS metric and an arbitrary $z$-dependent dilaton. In Sec. III we discuss the method we use to obtain a family of strictly isospectral potentials in the context of SUSY QM, and we explain how we used this to obtain a differential equation that allowed us to extract a family of metrics and/or dilatons. As an example, we present in detail the scalar glueball case with an AdS metric, and we obtain a family of dilatons. In Sec. IV we discuss the main problem in AdS/QCD models with chiral symmetry breaking. We also discuss some limits for dilatons in AdS spaces and some limits for metrics in the case that quadratic dilatons are used in AdS / QCD models, incorporating chiral symmetry breaking in a proper way. In Sec. V we present the possibility for using SUSY QM techniques to obtain families of dilatons and metrics consistent with the guidelines proposed; at the end of that section, we introduce a brief discussion about the interpolation functions for dilatons. Finally, in Sec. VI we present our conclusions and some future perspectives for additional work in this area.




\section{Scalar fields}

We consider a SW model, where action for scalar fields in spaces with 5D with dilaton is \cite{Colangelo:2008us, Vega:2008af, Gutsche:2011vb, dePaula:2009za}

\begin{equation}
\label{Accion}
S = - \int \mathit{d}^{5} x \sqrt{-g} e^{-\phi (z)} \frac{1}{2}[g^{MN}\partial
_{M} \Phi \partial _{N} \Phi + m^{2}_{5} \Phi ^{2}],
\end{equation}
with $M,N = 0, 1, 2, 3, z$, where $z$ is a holographic coordinate, $\phi (z)$ is a scalar dilaton field (used to break the conformal invariance in SW models) and $m_{5}$ is the mass of AdS modes along bulk [which is related to the dimension of operators that creates hadrons in field theory according to $m_{5}^2=\Delta(\Delta-4)$].

The metric considered was
\begin{equation}
\label{Metrica}
ds^{2} = a^{2}(z) \eta_{MN} dx^{M} dx^{N},
\end{equation}
where $\eta_{MN} = diag(-1,+1,+1,+1,+1)$, and $a^{2}(z)$ is a warp factor; therefore, in general we can consider AdS spaces asymptotically.

We considered scalar modes according to
\begin{equation}
\label{Modos}
\Phi(x) = e^{- i \mathcal{P} \cdot \mathcal{X}} f(z),
\end{equation}
where $\mathcal{P}$ and $\mathcal{X}$ correspond to momentum and position in boundary 4D and $\mathcal{P}^{2} = M^{2}$, where $M$ is the mass of hadron studied with this model.

Starting from equation of motion obtained from (\ref{Accion}), and by using (\ref{Modos}), it was possible to get an equation to $f(z)$
\begin{equation}
\label{EOM}
- f''(z) + \biggr( \phi'(z) - 3 \frac{a'(z)}{a(z)} \biggr) f'(z) + a^{2}(z) m_{5}^{2} f(z) = M^{2} f(z),
\end{equation}
which, by transforming
\begin{equation}
\label{Transformacion}
f(z)=\exp \biggr[\frac{1}{2} (\phi(z) - 3 ln~a(z)) \biggl] \psi (z),
\end{equation}
gives rise to a Schr\"odinger-type equation of the form
\begin{equation}
\label{EcSchro}
[ -\partial _{z}^{2}+V(z)] \psi (z) = M^{2} \psi (z),
\end{equation}
where $V(z)$ is the potential
\begin{widetext}
\begin{equation}
\label{Potencial}
V(z)= m_{5}^{2} a^{2}(z) + \frac{3}{4} \biggr( \frac{a'(z)}{a(z)} \biggl)^{2} - \frac{3}{2} \biggr( \frac{a'(z)}{a(z)} \biggl) \phi'(z) + \frac{1}{4} \phi'^{2}(z) + \frac{3}{2} \frac{a''(z)}{a(z)} - \frac{1}{2} \phi''(z).
\end{equation}
\end{widetext}

Through the use of $a(z) = R/z$ (where $R$ is the AdS radius) and $\phi (z)=\kappa ^{2}z^{2}$, the conventional AdS/QCD model was obtained \cite{Colangelo:2008us, Vega:2008af, Gutsche:2011vb}, which considers an AdS metric and a quadratic dilaton, and in this case, the potential is 
\begin{equation}
\label{PotencialSW}
V(z)=\frac{1}{z^{2}} \biggl(\frac{15}{4}+m_{5}^{2} R^{2} \biggr)+z^{2}\kappa ^{4}+2\kappa ^{2}.
\end{equation}

For this case, a solution to the eigenvalue problem (\ref{EcSchro}) is
\begin{equation}
\label{Sol}
\psi (z) = N_{n} \exp ^{-\frac{1}{2}z^{2}\kappa ^{2}} z^{(\frac{1}{2}+\sqrt{
4+m_{5}^{2} R^{2}})} L_{n}^{\sqrt{4+m_{5}^{2} R^{2}}}(z^{2}\kappa ^{2}),
\end{equation}
where $N_{n}$ is a normalization constant, that for the special case $n=0$ (that will be used later), is
\begin{equation}
N_{0} = \frac{2 \kappa^{2(1+\sqrt{
4+m_{5}^{2} R^{2}})}}{\Gamma(1+\sqrt{
4+m_{5}^{2} R^{2}})}. \nonumber
\end{equation}

The mass spectrum is given by
\begin{equation}
\label{Masas}
M^{2}_{n} = 4 \kappa ^{2} \biggr( n + \frac{1}{2} \sqrt{4+m_{5}^{2} R^{2}} + 1 \biggl).
\end{equation}

As can be seen in the previous expression, SW models with an AdS metric and quadratic dilaton are in agreement with Regge trajectories because $M^{2}_{n}~vs~n$ is linear.




\begin{figure}[t]
\includegraphics[width=3.3 in]{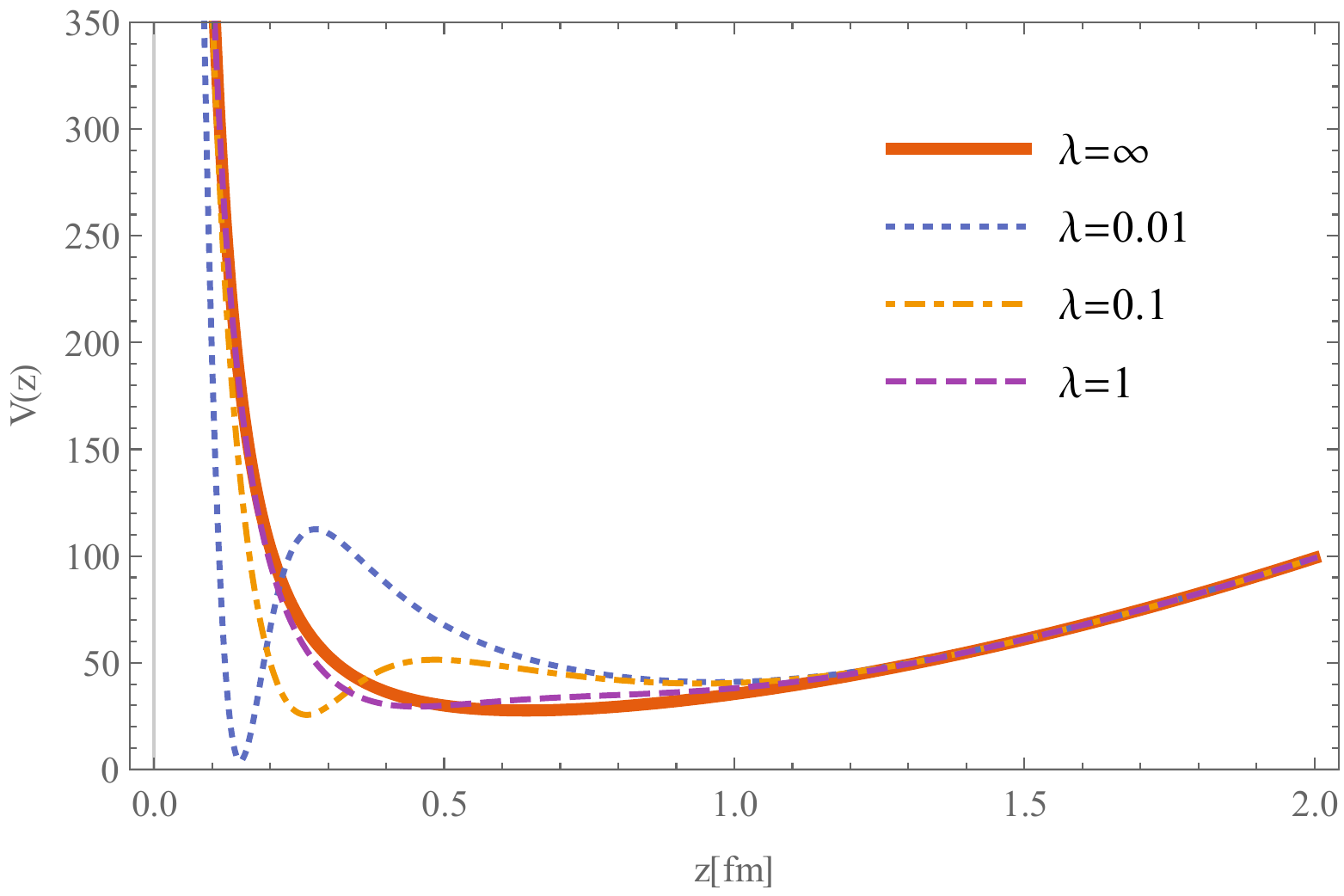}
\caption{Some potentials of isospectral families with different values for $\lambda$. When $\lambda= \infty $ (\ref{PotencialSW}) it recovers the original potential (\ref{Isopotencial}). Values used are $m_{5}=0$ and $\kappa = 428~MeV$.}
\label{fig:Isopotenciales}
\end{figure}

\section{Family of isospectral dilatons and metrics considerings SUSY QM techniques}

As we saw in the case of SW models, if we start with an AdS metric and a quadratic dilaton it is possible to obtain a Schr\"odinger-type equation associated with different hadrons on the AdS side. Although not a common practice, using the reverse (i.e., starting from a given potential) it is possible to extract a metric and a dilaton that reproduce this potential. This is the base of procedure discussed in this section.

For Schr\"odinger-type equations, there are several procedures that make it possible to obtain a family of strictly  isospectral potentials associated with a given potential \cite{AbrahamMoses, Luban:1986iv, Pursey:1986nc, Khare:1989ki, Cooper:1994eh}. As we show in this paper, this technique can be used to get families of metrics and/or dilatons for SW models for a given metric and dilaton. From among the different isospectral transformations, in this paper we used a procedure discussed in \cite{Cooper:1994eh}, that considers SUSY QM to obtain a family of strictly isospectral potentials depending on a parameter.

According to \cite{Cooper:1994eh}, two potentials $V_{1}$ and $\widehat{V}_{1}$ are strictly isospectral if
\begin{equation}
\label{Isopotencial}
\widehat{V}_{1}(z)=V_{1}(z)-2\frac{d^{2}}{dz^{2}}\ln [\emph{I}(z)+\lambda ].
\end{equation}
In this case, $V_{1}(z)$ is the original potential (\ref{PotencialSW}), $\lambda$ is and arbitrary parameter and the function $I(z)$, for coordinate between zero and infinity, is given by
\[
I(z) = \int_{0}^{z} \psi_{1}^{2}(z') dz',
\]
where $\psi_{1}$ corresponds to the ground state of $V_{1}$, in our case is given by (\ref{Sol}) with $n = 0$; using $R=1$ we obtained
\begin{equation}
I(z) =1-\frac{\Gamma (1+\sqrt{4+m_{5}^{2}} ,z^{2}\kappa ^{2})}{\Gamma (1+\sqrt{4+m_{5}^{2}})}.
\end{equation}

Figure (\ref{fig:Isopotenciales}) shows several strictly isospectral potential that appeared from (\ref{Isopotencial}), starting with $V_{1}(z)$ given by (\ref{PotencialSW}) for scalar glueballs ($m_{5} = 0$); it is possible to compare different potentials in this family in relation to the original one, and note also that changes in potentials are produced in a region close to the origin in $z$ and depends on $\lambda$. At higher values of $\lambda$, changes in family potentials are less, so when $\lambda \rightarrow \infty$ we recover the original potential. Again, all potentials in Fig. 1 produce the same spectrum for scalar glueballs, which in this case it is Regge-like.

Each member of this family was obtained by using (\ref{Isopotencial}) which could be considered as having been generated for a new metric and/or a new dilaton. For this reason we could match (\ref{Isopotencial}) and (\ref{Potencial}) and get
\begin{widetext}
\begin{equation}
\label{EcuacionFamilia}
V_{1}(z)-2\frac{d^{2}}{dz^{2}}\ln [\emph{I}(z)+\lambda ] = m_{5}^{2} a^{2}(z) + \frac{3}{4} \biggr( \frac{a'(z)}{a(z)} \biggl)^{2} - \frac{3}{2} \biggr( \frac{a'(z)}{a(z)} \biggl) \phi'(z) + \frac{1}{4} \phi'^{2}(z) + \frac{3}{2} \frac{a''(z)}{a(z)} - \frac{1}{2} \phi''(z).
\end{equation}

Notice that if in the last expression we use a specified metric, we get a differential equation to the family of dilatons. In this case we considered an AdS metrics, and the equation for dilaton is

\begin{equation}
\label{EcuacionFamiliaDilaton}
V_{1}(z)-2\frac{d^{2}}{dz^{2}}\ln [\emph{I}(z)+\lambda ] = \frac{15}{4 z^{2}} + \frac{m_{5}^{2} R^{2}}{z^{2}} + \frac{3 \phi'(z)}{2 z} + \frac{1}{4} \phi'^{2}(z) - \frac{\phi''(z)}{2}.
\end{equation}

On the other hand, by using Eq.(\ref{EcuacionFamilia}) for a definite dilaton can obtain a familly of metrics. By considering a quadratic dilaton, the equation for family of metrics is
\begin{equation}
\label{EcuacionFamiliaMetricas}
V_{1}(z)-2\frac{d^{2}}{dz^{2}}\ln [\emph{I}(z)+\lambda ] = -\kappa^{2} + \kappa^{4} z^{2} + m_{5}^{2} R^{2} a^{2}(z) - 3 \kappa^{2} z \frac{a'(z)}{a(z)} + \frac{3}{4} \biggr( \frac{a'(z)}{a(z)} \biggr)^{2} + \frac{3}{2} \frac{a''(z)}{a(z)}.
\end{equation}
\end{widetext}

\begin{figure}[t]
\includegraphics[width=3.3 in]{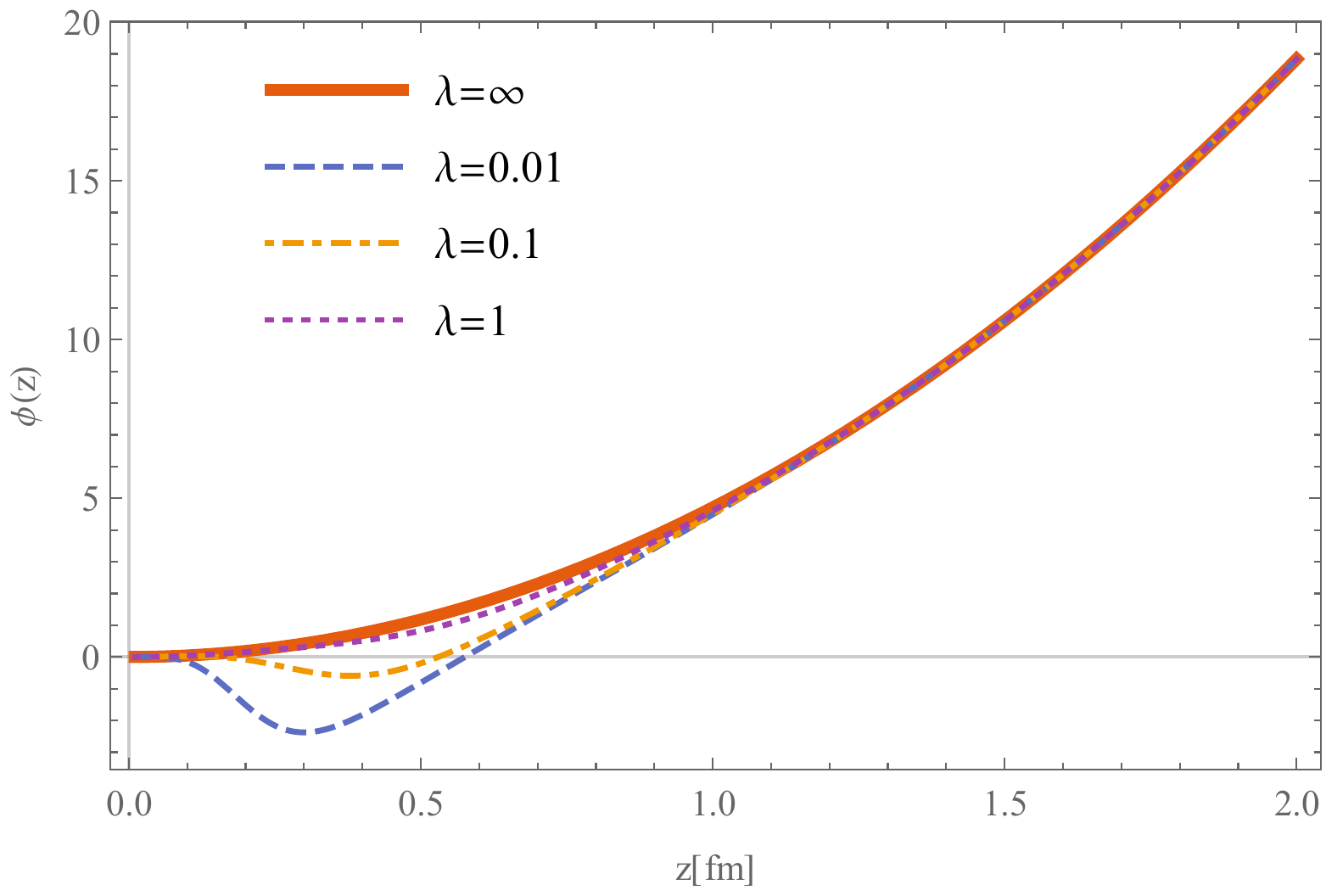}
\caption{Some isospectrals dilaton associated with potentials shown in fig. (\ref{fig:Isopotenciales}), with $m_{5}=0$ and $\kappa = 428~MeV$.}
\label{fig:Isodilatones}
\end{figure}

Figure (\ref{fig:Isodilatones}) shows several dilatons of family obtained solving (\ref{EcuacionFamiliaDilaton}).

It is important to verify that this procedure produces potentials with the same behaviour at low and high $z$ as the original one. However, as we show in next section, this must be changed in order to obtain dilatons or metrics appropriated in AdS/QCD models. This must be considered when using boundary conditions to solve the differential equations suggested in this section, and it obligates us to introduce modifications on them.




\section{Improvements in AdS/QCD models with chiral symmetry breaking}

Although the spectrum in bottom-up models with an AdS metric and quadratic dilaton is well 
reproduced in the mesonic sector \cite{deTeramond:2014asa, Dosch:2015nwa, Dosch:2015bca, Brodsky:2015oia, Branz:2010ub, Vega:2008te}, the 
incorporation of chiral symmetry breaking in this approach is far from QCD and needs improvements. This occurs because the scalar field in the bulk is dual to bilinear operator $q \bar{q}$, whose vacuum expectation value (VEV) ---which is responsible for chiral symmetry breaking in this model--- does not allow its explicit and spontaneous breaking in an independent and simultaneous 
way \cite{Colangelo:2008us, Gherghetta:2009ac, Kwee:2007nq, Vega:2010ne, Vega:2011tg}. The origin of this problem can be found in the solution of the VEV equation, which produces a finite action at the limit $z \rightarrow \infty$, given by

\begin{equation}
\label{vevSW}
v(z) \sim m_{q} z U(1/2,0,z^{2}),
\end{equation}
where $U(a,b,z)$ is a hypergeometric confluent function. According to the AdS/QCD dictionary established in \cite{Erlich:2005qh}, the VEV limit $z \rightarrow 0$ must be
\begin{equation}
\label{vevZ0}
v(z) \sim \alpha z + \beta z^{3}.
\end{equation}
In these models, quark mass $m_{q}$ and chiral condensate $\sigma$ are related with a constant that appears in (\ref{vevZ0}) through \cite{Erlich:2005qh}
\begin{equation}
\label{Paramvev}
m_{q} = \frac{\alpha R}{\zeta}~~~~and~~~~\sigma = \beta R \zeta,
\end{equation}
where $\zeta$ is a normalization parameter introduced in \cite{Cherman:2008eh}.

According to Gell-Mann--Oakes--Renner (GMOR) relationship, $m_{q}$ and $\sigma$ are related by  $m_{\pi}^{2} f_{\pi}^{2} = 2 m_{q} \sigma$; if we expand (\ref{vevSW}) to small values of the holographic coordinate, we get $\alpha \sim m_{q}$ and $\beta \sim \sigma \sim m_{q}$. This means that when quark mass is zero in this model then the chiral condensate is also zero, and this is not in agreement with GMOR.

The literature has some alternatives to solve this problem. For example, a negative quadratic dilaton could be a possible alternative \cite{Zuo:2009dz}, but it predicts an extra massless mode that are not observed \cite{Karch:2010eg}. It was recently suggested \cite{Chelabi:2015cwn} that an interpolation between two quadratic dilatons, one negative to small z and another positive to high z, could solve the problem without spurious modes that appear only with a negative dilaton. Another way to improve this model was studied in \cite{Gherghetta:2009ac}, where authors considered adding a quartic term in the scalar potential; in this case, they obtained
\begin{equation}
\label{Condicion}
\phi' = \frac{1}{a^{3} v'} \biggl[ \partial_{z} (a^{3} v') - a^{5} \biggl( m_{X}^{2} v - \frac{\kappa}{2} v^{3} \biggr) \biggr]
\end{equation}
which corresponds to an equation that relates the dilaton field $\phi(z)$, the warp factor in metric $a(z)$, VEV $v(z)$ and a constant $\kappa$ introduced in quartic term in action.

We can properly incorporate chiral symmetry breaking in soft-wall models by using (\ref{Condicion}) with an AdS metric [$a(z) = R/z$] and
\begin{equation}
\label{vev}
v(z)  = \frac{z}{R} (A + B~tanh~Cz^{2}),
\end{equation}
which is an interpolation between the limits to small and high values in $z$ for VEV. In this case dilaton satisfies

\begin{equation}
\label{ExtremosGherghetta1}
\phi (z \rightarrow 0) = \kappa_{0}^{2} z^{2},
\end{equation}
\begin{equation}
\label{ExtremosGherghetta2}
\phi (z \rightarrow \infty) = \kappa_{\infty}^{2} z^{2},
\end{equation}
where $\kappa_{0}^{2} = - \kappa_{\infty}^{2}$ in \cite{Chelabi:2015cwn} and $\kappa_{0}^{2} \neq \kappa_{\infty}^{2}$ in \cite{Gherghetta:2009ac}. We notice that, without details, which dilatons corresponding to interpolations between limit (\ref{ExtremosGherghetta1}) and (\ref{ExtremosGherghetta2}) can, according to some authors, improve other facts in the SW model (e.g., \cite{Li:2013oda}).

On the other hand, for a given dilaton, Eq.(\ref{Condicion}) can be used to obtain the asymptotic behavior in the factor $a(z)$. In this work we consider a quadratic dilaton, and we find  
\begin{equation}
\label{ExtremosMetrica1}
a (z \rightarrow 0) = A_{0} + B_{0} z^{2},
\end{equation}
\begin{equation}
\label{ExtremosMetrica2}
a (z \rightarrow \infty) = A_{\infty} + B_{\infty}^{2} z^{2}.
\end{equation}
Notice that the factor $a(z)$ related to the quadratic dilaton is not asymptotically AdS, and this shape remains even in $\kappa = 0$ in (\ref{Condicion}); i.e., it is not necessary to have a quartic term in the action in this case. However, at the moment, there are not good reasons to seriously consider a space that is not asymptotically AdS.




\section{Family of dilatons and metrics for AdS/QCD models using SUSY QM techniques}

Dilatons in agreement with (\ref{ExtremosGherghetta1}) and (\ref{ExtremosGherghetta2}) cannot be obtained from (\ref{EcuacionFamiliaDilaton}), because the original potential and the strictly isospectral potential has the same behaviour at low and high z values; i.e., dilatons in family found with (\ref{EcuacionFamiliaDilaton}) have $\phi (z \rightarrow 0) = \phi (z \rightarrow \infty) = \kappa^{2} z^{2}$. However, it is possible to introduce small changes to (\ref{EcuacionFamiliaDilaton}) in order to obtain dilatons in agreement with (\ref{ExtremosGherghetta1}) and (\ref{ExtremosGherghetta2}). To achieve this, we added a new term in (\ref{EcuacionFamiliaDilaton}) 
that slightly modifies the behavior at small $z$.

We have two possible locations to introduce the change: on the right or left side in (\ref{EcuacionFamiliaDilaton}). Thus, instead of (\ref{EcuacionFamiliaDilaton}) we suggest:

\begin{widetext}

\begin{equation}
\label{ModificacionIzquierda}
\frac{1}{z^{2}} \biggr(\frac{15}{4}+m_{5}^{2} \biggl)e^{-\delta z^2}+z^{2}\kappa ^{4}+2\kappa ^{2}-2\frac{d^{2}}{dz^{2}}\ln [\emph{I}(z)+\lambda ] = \frac{1}{z^{2}} \biggr(\frac{15}{4}+m_{5}^{2} \biggl) + \frac{3 \phi'(z)}{2 z} + \frac{1}{4} \phi'^{2}(z) - \frac{\phi''(z)}{2}
\end{equation}
\begin{center}
or
\end{center}
\begin{equation}
\label{ModificacionDerecha}
\frac{1}{z^{2}} \biggr(\frac{15}{4}+m_{5}^{2} \biggl)+z^{2}\kappa ^{4}+2\kappa ^{2}-2\frac{d^{2}}{dz^{2}}\ln [\emph{I}(z)+\lambda ] = \frac{1}{z^{2}} \biggr( \frac{15}{4}+m_{5}^{2} \biggl) e^{-\delta z^2} + \frac{3 \phi'(z)}{2 z} + \frac{1}{4} \phi'^{2}(z) - \frac{\phi''(z)}{2}.
\end{equation}
Notice that both equations can be summarized in
\begin{equation}
\label{Modificacion}
(-1)^p[e^{-\delta z^2}-1]\frac{1}{z^{2}} \biggr(\frac{15}{4}+m_{5}^{2} \biggl)+z^{2}\kappa ^{4}+2\kappa ^{2}-2\frac{d^{2}}{dz^{2}}\ln [\emph{I}(z)+\lambda ] = \frac{3 \phi'(z)}{2 z} + \frac{1}{4} \phi'^{2}(z) - \frac{\phi''(z)}{2},
\end{equation}

\end{widetext}
where $\delta$ and $p$ are two parameters that modify the behavior of $\phi(z)$ at low $z$.

\begin{figure}[b]
\label{fig:Npositivo}
\includegraphics[width=3.3 in]{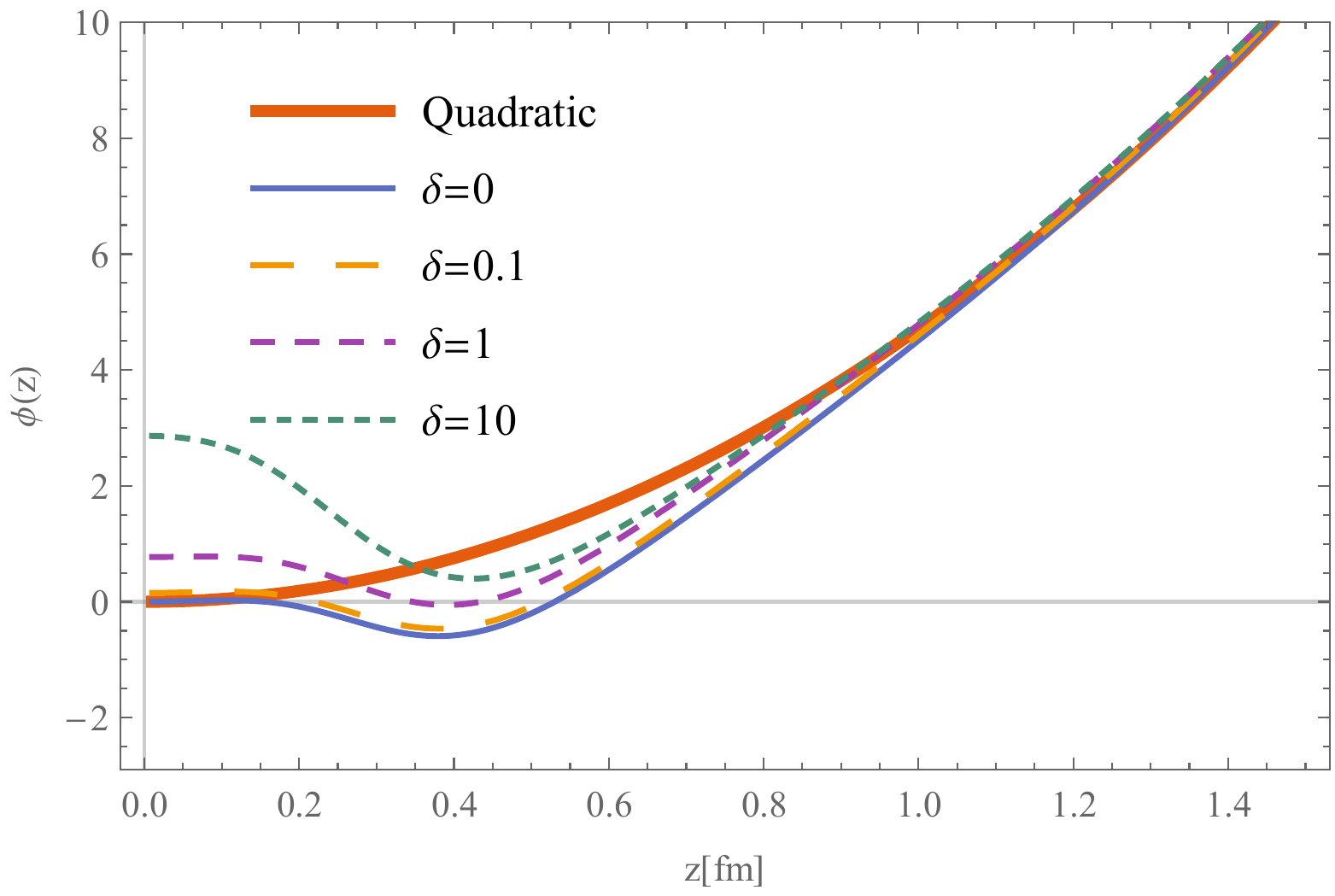}
\caption{Dilatons for different values of $\delta$, $p=even$, $\kappa = 428~MeV$, $m_{5}=0$ and $\lambda=0.1$}
\end{figure}

In Figs. 3 and 4 it is possible to see how $\delta$ modifies the magnitude of $\phi(z \rightarrow 0)$, and by using even or odd p values it was possible to choose the sign of $\phi(0)$ according to (\ref{ModificacionIzquierda}) or (\ref{ModificacionDerecha}). Increasing $\delta$ produces differences between the  spectrum calculated with quadratic dilaton and the one obtained from (\ref{ModificacionIzquierda}), (\ref{ModificacionDerecha}) or (\ref{Modificacion}). This constrains us to use small values for $\delta$ if we wish to obtain a dilaton that produces a spectrum very close to the original one. The behavior of the dilaton obtained in both extensions is $\phi \rightarrow C \pm \kappa_{0}^2 z^2$, with $C$ a constant.

It is worth noting in relation to implementation of chiral symmetry breaking in AdS/QCD models that the constant term is not important, and it can be dropped to calculate other quantities. In addition, whether the change suggested was implemented according to (\ref{ModificacionIzquierda}) or (\ref{ModificacionDerecha}) determines if is $\kappa_{0} < \kappa_{\infty}$ or $\kappa_{0} > \kappa_{\infty}$.

As noted earlier, a characteristic of a family of potentials obtained from SUSY QM techniques is that change occurs in a small region as compared to the original one, but the behaviour to small and large $z$ values is the same. This means that if we consider our procedure for obtaining a family of metrics related with a quadratic dilaton, each one will be asymptotically AdS.

An alternative procedure to built dilatons or metrics could be to consider interpolation functions between known limits. For example, for dilatons in models with AdS metrics it is possible to consider (\ref{ExtremosGherghetta1}) and (\ref{ExtremosGherghetta2}) as known limits, and in this case a possible dilaton could be

\begin{equation}
\label{DilatonInterpolado}
\phi(z) = A_{0} z^{2} + (A_{1} + B_{1}) z^{2}~ tanh(C~z^{2}).
\end{equation}

In this case, numerical calculation shows that this interpolation function produces a potential with Regge behaviour in a mass spectrum calculated with (\ref{EcSchro}).




\section{Summary and conclusions}

In this paper we discuss some alternatives for introducing dilatons and / or metrics for AdS / QCD models, considering as guideline that these models must reproduce mesonic spectrum and correctly implement chiral symmetry breaking. 

The main method discussed in this paper is based on SUSY QM techniques. This ensures that dilatons and 
metrics obtained with this procedure produce a proper mesonic spectrum. When we include as an additional condition the correct chiral symmetry breaking implementation, we found that in general this is not possible; however, introducing small changes in equations opens a possibility for obtaining dilatons that works in AdS spaces, but does not work to obtain metrics with quadratic dilatons.

Another alternative that we studied considers dilatons and/or metrics that correspond to interpolating 
functions between two limits. In our case we consider limits when $z \rightarrow 0$ and $z \rightarrow \infty$, and following Ref.~\cite{Gherghetta:2009ac}, the limits suggested in (\ref{ExtremosGherghetta1}) and (\ref{ExtremosGherghetta2}) for dilatons. We generalize the idea to obtain limits for $a(z)$ that define metrics. We found that in the case of quadratic dilatons the metric cannot be asymptotically AdS 
if chiral symmetry breaking is intended to be implemented in AdS/QCD models. This drawback means that quadratic dilatons are not a good choice in models with AdS metrics.

In our opinion, dilatons presented here, or other that could be obtained through the procedures discussed in this paper, could be interestingly used to calculate hadronic properties in other uses of AdS/QCD models. These possibilities will be explored in future work.

\begin{figure}[b]
\label{fig:Nnegativo}
\includegraphics[width=3.3 in]{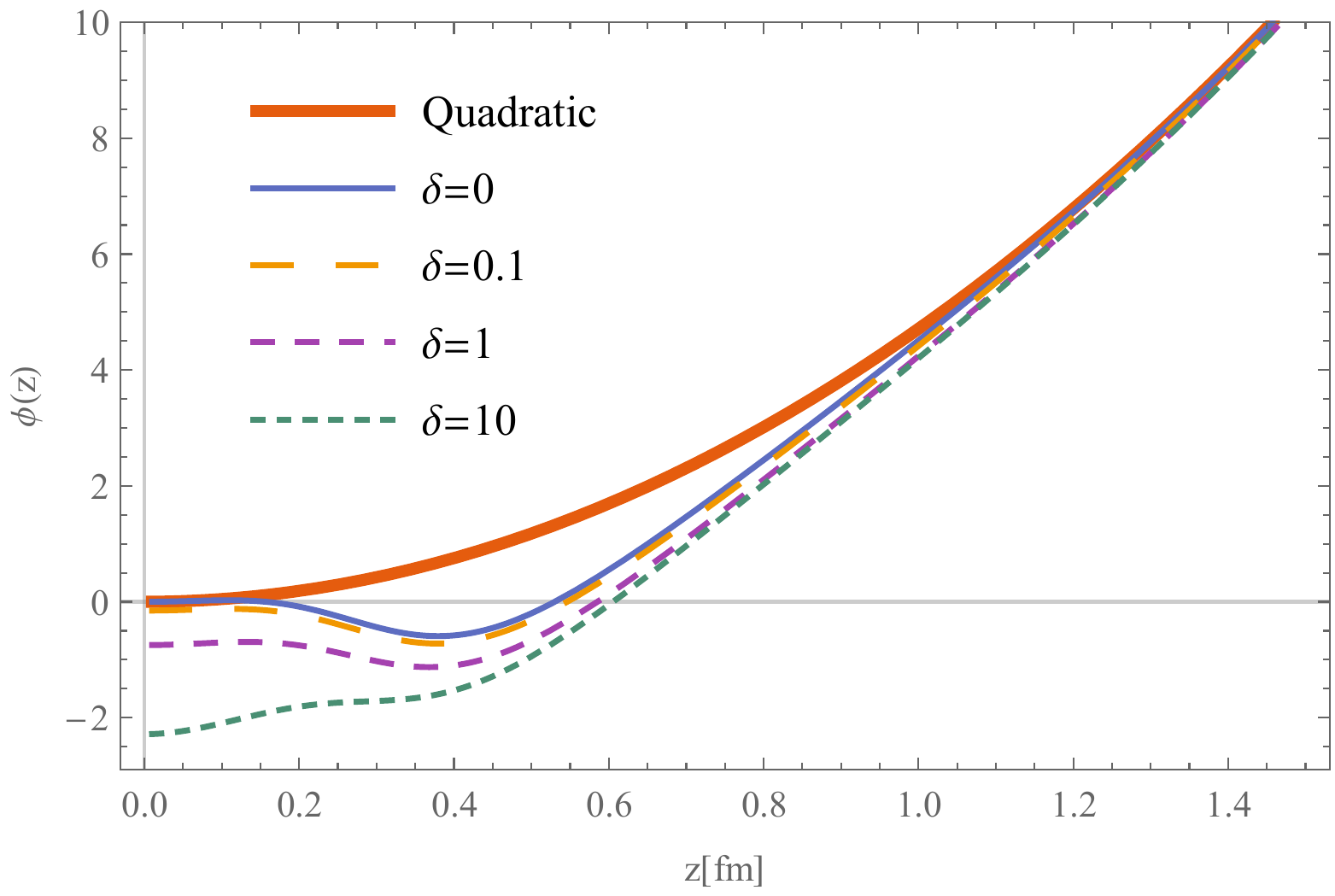}
\caption{Dilatons for different values of $\delta$, $p=odd$, $\kappa = 428~MeV$, $m_{5}=0$ and $\lambda=0.1$ }
\end{figure}




\begin{acknowledgments}

The authors thank Valery E. Lyubovitskij for a critical reading the manuscript and useful remarks. This work was supported by FONDECYT (Chile) under Grant No. 1141280 and by CONICYT (Chile) under Grant No. 7912010025.

\end{acknowledgments}

\end{document}